\documentstyle[11pt,newpasp,epsf,twoside]{article}
\newcommand{\ha}{H$\alpha$}
\newcommand{\msun}{M$_{\odot}$}
\newcommand{\reff}{\reference}

\begin{document}
\title{Do Herbig Ae/Be stars have discs?}
 \author{Jorick S. Vink$^1$, Janet E. Drew$^1$, Tim J. Harries$^2$, Ren\'e D. Oudmaijer$^3$}
\affil{
$^1$Imperial College of Science, Technology and Medicine,\\
~Blackett Laboratory, Prince Consort Road, London, SW7 2BZ, UK\\   
$^2$ School of Physics, University of Exeter, Stocker Road, \\
~Exeter EX4 4QL, UK\\
$^3$ The Department of Physics and Astronomy, E C Stoner Building,\\
~Leeds, LS2 9JT, UK}

\begin{abstract}
\ha\ spectropolarimetry on Herbig Ae/Be stars shows that 
the innermost regions of intermediate mass (2 -- 15 \msun) 
Pre-Main Sequence stars are flattened. This may be 
the best evidence to date that the higher mass Herbig Be 
stars are embedded in circumstellar discs. 
A second outcome of our study is that the spectropolarimetric 
signatures for the lower mass Herbig Ae stars differ from those of the 
higher mass Herbig Be stars. 
Depolarisations across \ha\ are observed in the Herbig Be group, 
whereas line polarisations are common amongst the Herbig Ae 
stars in our sample. 
These line polarisation effects can be understood in terms of 
a compact \ha\ source that is polarised by a rotating 
disc-like configuration. The difference we detect between 
the Herbig Be and Ae stars may be the first indication that there 
is a transition in the Hertzsprung-Russell Diagram from magnetic 
accretion at spectral type A to disc accretion at spectral type B.
However, it is also possible that the compact polarised line component, 
present in the Herbig Ae stars, is masked in the Herbig Be stars 
due to their higher levels of \ha\ emission.
\end{abstract}

\section{Introduction}

Herbig Ae/Be stars are intermediate mass (2 -- 15 \msun) 
Pre-Main Sequence stars. They are the most massive counterparts 
of the lower mass T~Tauri stars that are visible in the optical.
Therefore, Herbig stars are ideal objects to determine whether 
key aspects of the T~Tauri phase, such as discs and magnetic fields, are 
also of physical relevance to higher mass star formation. 
Yet, even the most basic question of whether Herbig Ae/Be stars are 
embedded in accretion discs has not been answered. 
Ideally, one would like to image the environments around Herbig Ae/Be stars
directly, but these attempts have so far resulted in contradictory 
results. For instance, CO observations by Mannings \& Sargent (2000) 
in the sub-millimetre regime show flattened structures on large 
scales (up to 1000 AU), but Millan-Gabet et al. (2001), who 
conducted a 2-telescope interferometric study at near-infrared 
wavelengths, concluded that accretion disc models can be ruled out on scales between 0.5 -- 5 
AU, with spherical models reproducing the visibility data much better. 
It is obvious from the above that there is a need for an 
independent approach to answer this problem, providing 
a diagnostic that can resolve circumstellar structures in the innermost 
regions around young stars. Spectropolarimetry across emission lines 
is just such a tool, as it can probe spatial scales on the order of stellar 
radii (i.e. 1/100s AUs) rather than AUs.

\section{The tool of linear spectropolarimetry}

The application of linear spectropolarimetry was first established in studies of 
classical Be stars. 
The method is based on the expectation that the H$\alpha$ 
photons -- formed over a large volume in a circumstellar 
medium -- undergo less electron scatterings off a disc 
than the stellar continuum photons do. Consequently, 
the line flux will be much less polarised than the continuum, and 
a change 
in the polarisation across the line occurs. We refer to this depolarisation 
as the classical line-effect because of its first appearance in observations 
of classical Be stars.
The high incidence of these depolarisations among such 
objects (26 out of 44 in Poeckert \& Marlborough 1976) indicated 
that Be star envelopes are not spherically symmetric. 
In its time, this evidence was taken as showing that classical Be stars possess 
discs -- a result later confirmed by interferometric imaging.
We use the tool of \ha\ spectropolarimetry on a large sample of bright Herbig stars. 
The sample needs to be large to make a proper 
distinction between intrinsic geometrical effects and our viewing angle. When viewed 
face-on, one would not expect to detect a line effect, due to the symmetry projected 
on the sky.
When viewed edge-on, the line effect is maximal. Our sample so far consists 
of 12 Herbig Be stars and 11 Herbig Ae stars (see Vink et al. 2002). 

\section{The Herbig Be stars}

A typical observation of the Herbig Be star BD+40 4124 is shown in Fig.~1.
Note the presence of the classical line-effect, This directly implies 
that for this object the electron-scattering region is not spherically symmetric,
i.e. it is flattened.  

The frequency of depolarisations detected in the HBe star sample (7/12; Oudmaijer \& Drew 
1999 + Vink et al. 2002) is particularly interesting because it is essentially 
the same as was found for classical Be stars (Poeckert \& Marlborough 1976).  
Continuing the analogy with classical Be stars, our H$\alpha$ spectropolarimetry indicates 
that the higher mass Herbig Be stars are embedded in 
flattened structures. Note however, that just the presence of these flattened 
structures is not evidence for continuing accretion. 

\begin{figure} 
\plotone{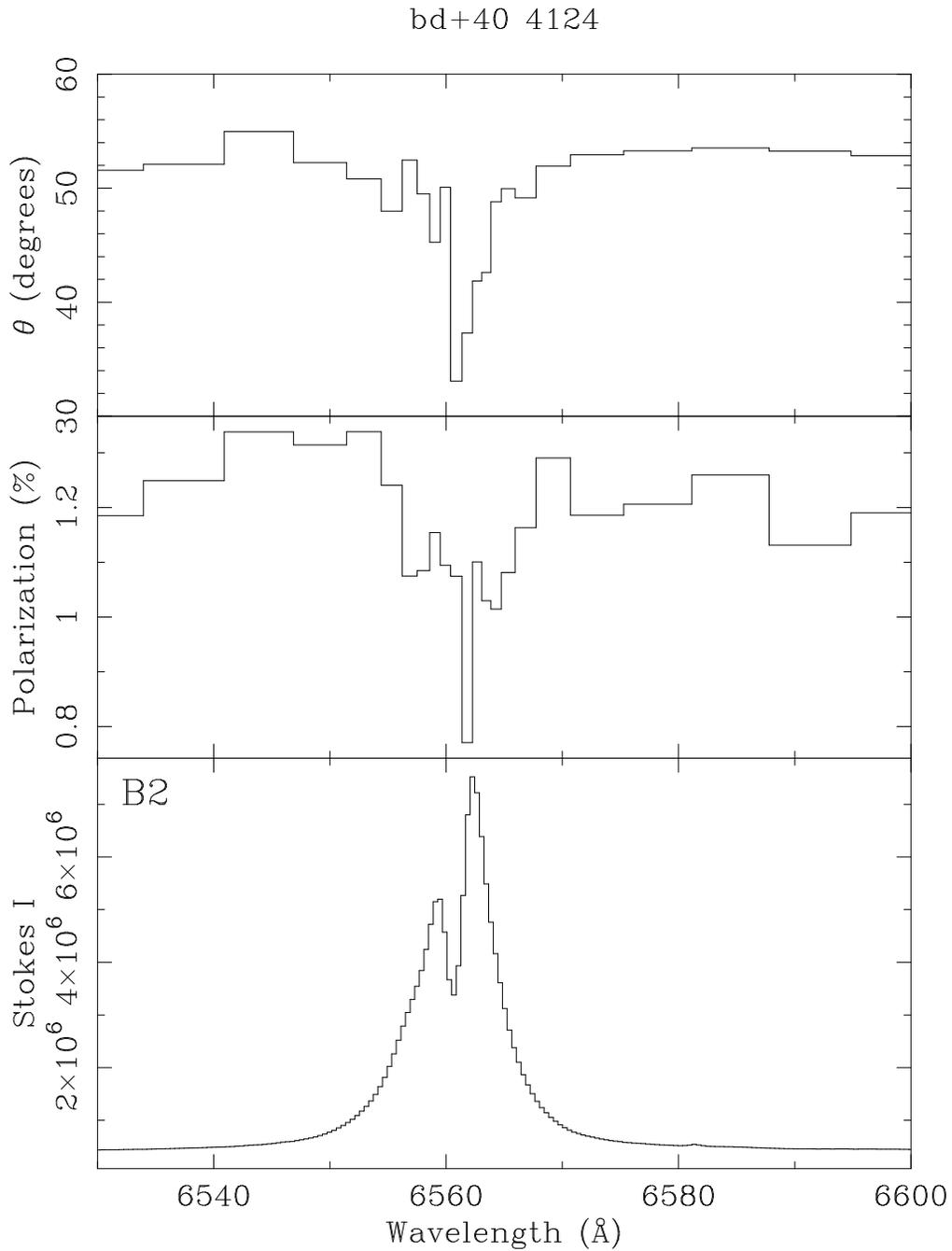}
\caption{The Herbig Be star BD+40~4124. The normal (intensity) spectrum is shown in the 
lower panel, the polarisation percentage is indicated in the middle panel, and the position 
angle (PA) is plotted in the upper panel. The data are binned to an error (1 {$\sigma$}) of  
0.05 \%. Note the depolarisation across the H{$\alpha$} emission line.}
\end{figure}

\begin{figure} 
\plotone{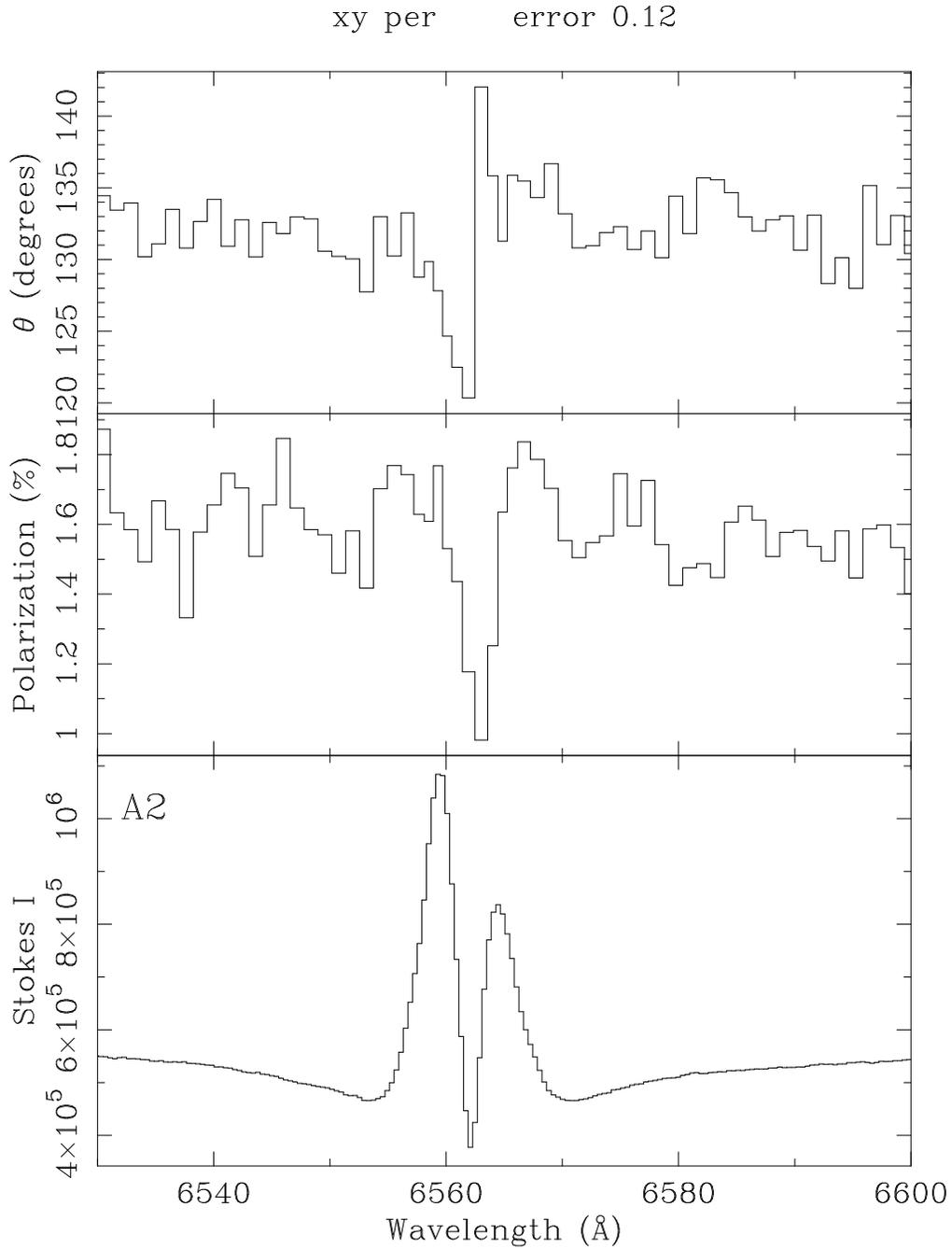}
\caption{The Herbig Ae star XY~Per. The panels are as in Fig.~1. The data are binned to an 
error of 0.12 \%. Note the rotation in the Position Angle which can be 
can be explained with a compact H{$\alpha$} source in a rotating circumstellar disc.}
\end{figure}

\section{The Herbig Ae stars}

At later spectral type we detect 
spectropolarimetry characteristics differing from the earlier type Herbig Be 
(HBe) stars. A switch in phenomenology may be expected to occur at some point 
working down the stellar mass range, as different physical mechanisms might play 
a more prominent role at different spectral types.  For instance, radiation 
pressure forces are likely to play a role for the higher luminosity stars at 
the early B types, whereas magnetic fields may become more dynamically 
prominent at the later A types. The magnetically-channelled accretion model 
that is commonly applied to the lowest mass Pre-Main Sequence T~Tauri stars 
may also be a 
suitable model as early as spectral type A.  If it does operate, the inner accretion disc 
around the star is truncated by the 
magnetic field, and the depolarisation effect may then be absent 
because the inner hole will necessarily lead to reduced intrinsic continuum 
polarisation.  Alternatively, the channelled accretion may produce a 
relatively bright and compact source of H$\alpha$ emission that may be 
scattered either within the accretion column itself or within the disrupted 
disc.  This in turn may yield a polarisation signature at H$\alpha$ 
that is more complex than the simple depolarisation effect.

A typical Herbig Ae star (XY~Per) is shown in Figure~2. Note first that   
the polarisation signature is more complicated than in Fig.~1, and 
that the data are clearly at variance with the depolarisation 
picture described above.
The observed ``flip'' in the Position Angle can be understood by 
the presence of a compact H$\alpha$ source, located within a rotating 
distribution of scatterers (see also Pontefract et al. 2000). 
The data for 9/11 Herbig stars in our sample imply the presence 
of a compact H$\alpha$ source. This could arise from continuous 
accretion on the stellar surface, perhaps 
magnetically-controlled, producing hot spots on the stellar surface. 
Future spectropolarimetric monitoring campaigns in conjunction with 
radiative transfer modelling (using {\sc torus}; Harries 2000) should 
be able to test this.

\section{Difference between Herbig Be and Ae stars?}

The spectropolarimetric differences between the Herbig Be and Ae stars (as seen 
in Figures 1 and 2) may be the first indication that there is a transition in 
the Hertzsprung-Russell Diagram from magnetic accretion at spectral type A to disc 
accretion at spectral type B. However, there are other 
differences between Herbig Ae and Be stars. Most notable, the \ha\ emission in 
Herbig Be stars is much more extended in volume and optical depth. It is therefore 
possible that the compact spectropolarimetric signature seen in the Ae stars 
may be produced in the Herbig Be stars also, but is completely masked there by \ha\ emission. 

To conclude, our data indicate that flattened structures around Herbig Ae/Be stars are 
common on the smallest scales, suggesting that the higher mass Pre-Main Sequence 
stars may well be embedded in accretion discs.

\end{document}